\documentclass[10pt,a4paper]{article}
\usepackage[utf8]{inputenc}
\usepackage{amsmath}
\usepackage{amsfonts}
\usepackage{amssymb}

\usepackage[electronic]{ifsym}
\usepackage{indentfirst,latexsym}
\usepackage{bm}
\usepackage{bbm}
\usepackage{mathrsfs}
\usepackage{amsmath,amsfonts,amsthm,amscd,amssymb}
\usepackage{pifont}
\usepackage{orcidlink}
\usepackage{url}
\usepackage{dirtytalk}

\usepackage{graphicx}
\usepackage[left=2cm,right=2cm,top=2cm,bottom=2cm]{geometry}

\usepackage{algorithm}         
\usepackage{algorithmicx}
\usepackage{algpseudocode}
\usepackage{float}
\usepackage{extarrows}

\newcommand{\Fig}{\textbf{Figure}~}
\newcommand{\Tab}{\textbf{Table}~}
\newcommand{\Thm}{\textbf{Theorem}~}
\newcommand{\Lem}{\textbf{Lemma}~}
\newcommand{\Cor}{\textbf{Corollary}~}
\newcommand{\App}{\textbf{Appendix}~}

\algnewcommand\algorithmicswitch{\textbf{switch}}
\algnewcommand\algorithmiccase{\textbf{case}}
\algnewcommand\algorithmicdefault{\textbf{default}}

\algdef{SE}[SWITCH]{Switch}{EndSwitch}[1]{\algorithmicswitch\ #1\ \algorithmicdo}{\algorithmicend\ \algorithmicswitch}%
\algdef{SE}[CASE]{Case}{EndCase}[1]{\algorithmiccase\ #1}{\algorithmicend\ \algorithmiccase}%
\algdef{SE}[DEFAULT]{Default}{EndDefault}[1]{\algorithmicdefault\ #1}{\algorithmicend\ \algorithmicdefault}%

\makeatletter
\renewcommand{\ALG@name}{Algorithm}

\renewcommand{\bf}[1]{\mathbf{#1}}

\newcommand{\plpt}[2]{\binom{#1}{#2}}

\newcommand{\mfloor}[1]{ \left\lfloor {#1} \right\rfloor }
\newcommand{\mceil}[1]{ \left\lceil {#1} \right\rceil }

\newcommand{\mat}[1]{\bm{#1}}

\newcommand{\set}[1]{\left\{ #1 \right\}}

\newcommand{\abs}[1]{\left| #1 \right|}

\newcommand{\trsp}[1]{{#1}^\textsf{T}}

\newcommand{\inv}[1]{#1^{-1}}

\newcommand{\ES}[3]{\mathbb{#1}^{{#2}\times {#3}}}     

\DeclareMathOperator{\GCD}{GCD}

\newcommand{\scrd}[2]{{#1}_{\mathrm{#2}}}
\newcommand{\scrud}[3]{{#1}^{\mathrm{#2}}_{\mathrm{#3}}}

\newtheorem{thm}{Theorem }
\newtheorem{cor}[thm]{Corollary}
\newtheorem{lem}[thm]{Lemma}

\newtheorem{res}[thm]{Result}

\author{Hong-Yan Zhang$^{1,2}$\orcidlink{0000-0002-4400-9133}, Haoting Liu$^{2,1}$, 
 Rui-Jia Lin$^3$, Yu Zhou$^1$\\
\begin{tabular}{lll}
\small{$1$}.~\small{\textit{School of Information Science and Technology, Hainan Normal University, Haikou 571158, China}}\\
\small{$2$}.~\small{\textit{School of Automation, University of Science and Technology
Beijing, Beijing 100083, China}}\\
\small{$3$}.~\small{\textit{Information Network and Data Center, Hainan Normal University, Haikou 571158, China}}
\end{tabular}
}
\title{A Correction for the Paper \say{Symplectic geometry mode decomposition and its application to rotating machinery \\compound fault diagnosis}\thanks{Corresponding author: Hong-Yan Zhang, e-mail: hongyan@hainnu.edu.cn}}

\begin{document}
\maketitle

\begin{abstract}
The \textit{symplectic geometry mode decomposition} (SGMD) is a powerful method for decomposing time series, which is based on the \textit{diagonal averaging principle} (DAP) inherited from the \textit{singular spectrum analysis} (SSA). Although the authors of SGMD method generalized the form of the trajectory matrix in SSA, the DAP is not updated simultaneously. In this work, we pointed out the limitations of the SGMD method and fixed the bugs with the pulling back theorem for computing the given component of time series from the corresponding component of trajectory matrix.   
\end{abstract}

\tableofcontents

\section{Introduction}

The \textit{symplectic geometry mode decomposition} (SGMD)  is originally proposed by Pan et al. in the paper \say{\textbf{Symplectic geometry mode decomposition and its application to rotating machinery compound fault diagnosis}} published in \textsc{Mechanical System and Signal Processing} \cite{Pan2019sgmd} for decomposing  time series\footnote{The time series is also named with time sequence.}. 
The SGMD depends on the \textit{diagonal averaging principle} (DAP) proposed by Vautard et al. \cite{Vautard1992SSA} in the classic paper about the \textit{singular spectrum analysis} (SSA).

In recent five years, there are active researches about the SGMD, such as 
Jin et al. \cite{Jin2019}, Zhang \cite{ZhangGY2022Esgmd}, Guo et al. \cite{GuoJC2022, GuoJC2023}, Chen et al. \cite{ChenYJ2023}, 
 Liu et al. \cite{Liu2024SpSparse}, Hao et al. \cite{Hao2024}, Zhan et al \cite{ZhanPM2024}, and Xin et al. \cite{XinG2025Csgmd}. 
Although the researches which  demonstrate the merits of the SGMD method are still increasing, there is \textit{no doubt about the key principle  behind the SGMD}. It should be noted that there are two limitations for the DAP in SSA  
\begin{itemize}
\item it just holds for the time delay $\tau = 1$ in the step of constructing the trajectory matrix for embedding;
\item it just works for the time series denoted by $X=\set{x[n]}^N_{n=1}$ but  fails for the time series denoted by $X=\set{x[n]}^{N-1}_{n=0}$  due to different structures of the trajectory matrix.
\end{itemize} 
Although the trajectory matrix is generalized by Pan et al. in SGMD, the DAP is not generalized. Consequently, the reconstruction principle is not correct if the time delay $\tau>1$ and time series is denoted by $X=\set{x[n]}^{N-1}_{n=0}$. In this paper, our purpose is to fix theses bugs with a general formula in which both the time delay $\tau$ and type flag $s$ are involved explicitly.

\section{Preliminaries}
\subsection{Notations}

The time series $X$ of length $N\in \mathbb{N}$ can be denoted by $x[n]$ for simplicity. There are two types of concrete representations for different computer programming languages:
\begin{itemize}
\item type-1 for the Fortran/MATLAB/Octave/...
      \begin{align*}
       X = \set{x[n]: 1\le n\le N}
         = \set{x[1], x[2], \cdots, x[N]}
      \end{align*}
\item type-0 for the C/C++/Python/Java/Rust/...
      \begin{align*}
      X = \set{x[n]: 0\le n\le N-1} 
        = \set{x[0], x[1], \cdots, x[N-1]}
      \end{align*}
\end{itemize}
It is trivial to find that the unified formula for these two types can be expressed by
\begin{equation}
X=\set{x[n]: s\le n\le N-1+s}=\set{x[n]}^{N-1+s}_{n=s}
\end{equation}
for $s\in \set{0,1}$. The set of time sequence with length $N$ is denoted by
\begin{equation}
\mathscr{X} = \set{X=\set{x[n]}^{N-1+s}_{n=s}: N\in \mathbb{N}, x[n]\in \mathbb{R}},
\end{equation}
which is called the \textit{sequence space}. 
Suppose $d$ is the dimension of immersion space and 
\begin{equation}
m \xlongequal{\tau\in \mathbb{N}} = N - (d-1)\tau
\end{equation}
is the number of $d$-dim signals in the immersion space for the time series $X$.   
In the work of SSA \cite{Vautard1992SSA}, the trajectory matrix of the time series is expressed by
\begin{equation}
\begin{aligned}
\scrd{\mat{X}}{ssa} 
&= \begin{bmatrix}
x[1] & x[2] & \cdots & x[m] \\
x[2] & x[3] & \cdots & x[m+1] \\
\vdots & \vdots & \ddots & \vdots \\
x[d] & x[d+1] & \cdots & x[N]
\end{bmatrix}
\end{aligned} 
\end{equation} 
where $m$ is a positive integer such that 
\begin{equation}
m \xlongequal{\tau=1} N- (d-1)\tau = N-d+1
\end{equation}
In the work of SGMD \cite{Pan2019sgmd}, Pan et al.  generalized the form of the trajectory matrix with a time delay $\tau\in \mathbb{N}$, which is written by
\begin{equation} \label{eq-M-pan}
\begin{aligned}
\scrd{\mat{X}}{sgmd} 
&= \begin{bmatrix}
x[1] & x[1+\tau] & \cdots & x[1+(d-1)\tau] \\
x[2] & x[2+\tau] & \cdots & x[2+(d-1)\tau] \\
\vdots & \vdots & \ddots & \vdots \\
x[m] & x[m+\tau] & \cdots & x[m+(d-1)\tau] \\
\end{bmatrix}
\end{aligned} 
\end{equation} 
Obviously, if $\tau=1$, then $\scrd{\mat{X}}{sgmd} $ is the transpose of $\scrd{\mat{X}}{ssa}$.
In order to unify the notations, the embedding mapping parameterized by the type $s$ and time delay $\tau$ is denoted by 
\begin{equation} 
\begin{aligned} 
\Phi_{s,\tau}: \mathscr{X}&\to \ES{R}{d}{m} \\
X&\mapsto \mat{M}
\end{aligned}
\end{equation}
such that 
\begin{equation} \label{eq-M-s-tau}
\begin{aligned}
		\mat{M}
	    &=  \Phi_{s,\tau}(X) 
	    = \left(M^{i+s}_{j+s}\right)_{d\times m} =\left(x[i\tau+j+s] \right)_{d\times m} \\
		&= \begin{bmatrix}
			x[s] & x[s+ 1]  & \cdots & x[s+m-1]  \\
			x[\tau+s] & x[\tau+s+1]  & \cdots & x[\tau +s+m-1] \\
			\vdots & \vdots   & \ddots & \vdots \\
			x[(d-1)\tau+s] & x[(d-1)\tau+s+1] & \cdots & x[(d-1)\tau+s+m-1]
		\end{bmatrix}
\end{aligned}
\end{equation}
for $0\le i \le d-1$ and $0 \le j \le m-1$ for the time series 
$X$ of type-$s$ with length $N$.  Intuitively, the inverse of the embedding mapping
\begin{equation}
\begin{aligned}
\inv{\Phi}_{s,\tau}: \ES{R}{d}{m} &\to \mathscr{X} \\
\mat{M} &\mapsto X
\end{aligned}
\end{equation}
 can be regarded as the \textit{pulling back mapping} since it pull back the trajectory matrix to the original time series.

Please note  there are two types of representation methods for the time sequece, which leads to different forms of the trajectory matrix. Particularly, for $s = 0$, we have the following trajectory matrix of type-0 
\begin{equation} \label{eq-X-traj-mat-0}
	\begin{split}
		\mat{M} 
		&=\Phi_{0,\tau}(X) = [\bf{x}_0, \bf{x}_1, \cdots, \bf{x}_{m-1}]
		=  \left(M^i_j\right)_{d\times m} =\left(x[i\tau+j] \right)_{d\times m}\\
		&= \begin{bmatrix}
			x[0] & x[1]  & \cdots & x[m-1] \\
			x[\tau] & x[\tau+1]  & \cdots & x[\tau +m-1] \\
			\vdots & \vdots   & \ddots & \vdots \\
			x[(d-1)\tau] & x[(d-1)\tau+1] & \cdots & x[(d-1)\tau+m-1]
		\end{bmatrix}
\end{split}
\end{equation}
for the row index $0\le i \le d-1$ and column index $0\le j\le m-1$ 
where $\bf{x}_j = \trsp{\begin{bmatrix}x[j], x[\tau+j], \cdots, x[(d-1)\tau+j]\end{bmatrix}}$  is a $d$-dim vector. It is easy to find that 
 \begin{itemize}
 \item $ M^0_0 = x[0]$ and $M^{d-1}_{m-1} = x[(d-1)\tau + m-1] = x[N-1]$;
 \item each of the $x[n]$ for $0\le n\le N-1$ has been embedded in the trajectory matrix;
 \item for the given $n$, the $x[n]$ appears at the position of $\plpt{x}{y}$ such that $\tau x + y = n$ for $0\le x\le d-1$ and  $0\le y\le m-1$.
\end{itemize}

Similarly, for $s = 1$, we have the trajectory matrix of type-1 as follows
\begin{equation} \label{eq-X-traj-mat-1}
	\begin{split}
		\mat{M} 
		&=\Phi_{1,\tau}(X) = [\bf{x}_1, \bf{x}_2, \cdots, \bf{x}_{m}]
		=  \left(M^i_j\right)_{d\times m} =\left(x[(i-1)\tau+j] \right)_{d\times m}\\
		&=\begin{bmatrix}
			x[1] & x[2]  & \cdots & x[m] \\
			x[\tau+1] & x[\tau+2]  & \cdots & x[\tau +m] \\
			\vdots & \vdots   & \ddots & \vdots \\
			x[(d-1)\tau+1] & x[(d-1)\tau+2] & \cdots & x[(d-1)\tau+m]
		\end{bmatrix}
\end{split}
\end{equation}
for $1\le i \le d$ and  $1\le j\le m$ 
where 
$$\bf{x}_{j} = \trsp{\begin{bmatrix}x[j], x[\tau+j], \cdots, x[(d-1)\tau+j]\end{bmatrix}}$$
  is a $d$-dim signal vector. It is easy to find that 
 \begin{itemize}
 \item $ M^1_1 = x[1]$ and $X^d_m = x[(d-1)\tau + m] = x[N]$;
 \item each of the $x[n]$ for $1\le n\le N$ has been embedded into the trajectory matrix;
 \item for given $n$, the $x[n]$ appears at the position of $\plpt{i}{j}$ such that $\tau (i-1) + j = n$ for $1\le i\le d$ and
 $1\le j\le m$.
\end{itemize}
It is easy to check that
\begin{equation}
\scrd{\mat{X}}{ssa} = \left. \mat{M}\right|_{s=1,\tau=1}, \quad 
\scrd{\mat{X}}{sgmd} = \left.\trsp{\mat{M}}\right|_{s=1}.
\end{equation}

Suppose the trajectory matrix can be decomposed into $r$ components $\set{\mat{Z}_1, \cdots, \mat{Z}_k, \cdots, \mat{Z}_r}$ such that
\begin{equation}
\mat{M} = \sum^r_{k=1} \mat{Z}_k,
\end{equation}
then the $k$-th component $\mat{Z}_k\in \ES{R}{d}{m}$ can be used to compute the $k$-th component
\begin{equation}
X^{(k)}=\set{x^{(k)}[n]}^{N-1+s}_{n=s}\in \mathscr{X}, \quad 1\le k\le r
\end{equation}
of the time series $X=\set{x[n]}^{N-1+s}_{n=s}$. According to Pan et al \cite{Pan2019sgmd}, the matrices $\set{\mat{Z}_1, \cdots, \mat{Z}_r}$ are computed based on symplectic orthogonal transform. By comparison, they are computed with singular value decomposition in the SSA method.

\subsection{Binary Diophantine Equation $\tau x + y = n+s\tau$ with Rectangular Constraint}

\begin{figure*}[htbp]
\centering
\includegraphics[width=0.7\textwidth]{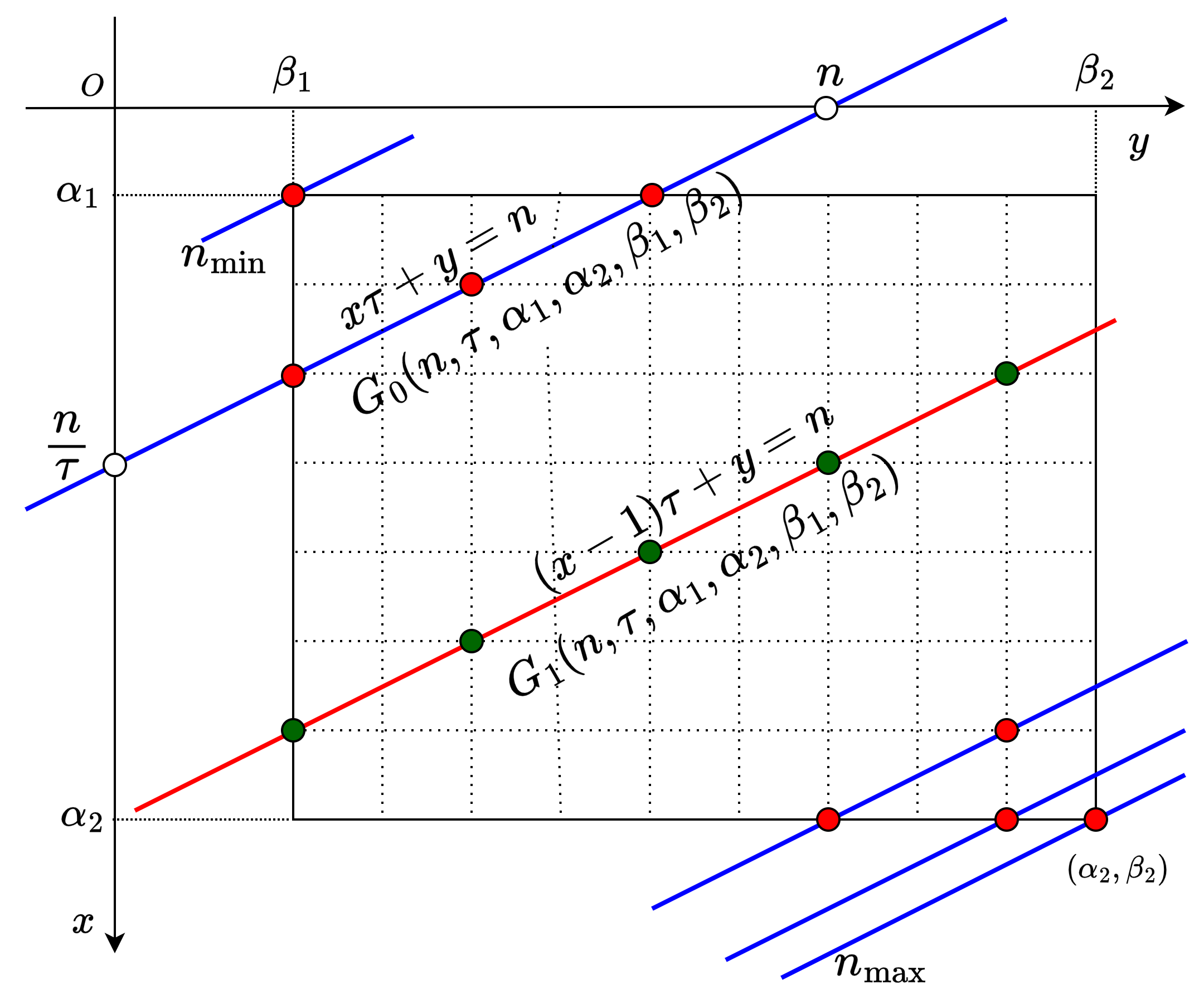} 
\caption{Geometric interpretation of $G_s(n,\tau,\alpha_1, \alpha_2,\beta_1,\beta_2)$} \label{fig-Dipahn-eq-geo}
\end{figure*}

Assume that $N, d, \tau\in \mathbb{N}$ are positive integers. Suppose that $\alpha_1, \alpha_2, \beta_1, \beta_2\in \mathbb{Z}^+$ such that $0\le \alpha_1 < \alpha_2$ and $0\le \beta_1<\beta_2$. Let
\begin{equation}
\begin{aligned}
\Omega = \Omega(\alpha_1, \alpha_2, \beta_1, \beta_2)
 =\set{\plpt{x}{y}\in \mathbb{Z}^2: \alpha_1 \le x \le \alpha_2, \beta_1\le y \le \beta_2}
\end{aligned}
\end{equation}
 For the given $n\in \mathbb{Z}^+$ and $\tau\in \mathbb{N}$,  we have the greatest common divisor $\GCD(\tau,1) = 1$ and $n> \tau \cdot 1 -\tau - 1 = -1$, thus the set of solutions to the Diophantine equation 
\begin{equation} \label{eq-xy-diop}
\tau x + y = n+s\tau, \quad \forall n\in \mathbb{Z}^+
\end{equation}
must be non-empty according to the \Thm  \ref{thm-appendix-1}  and \Thm  \ref{thm-appendix-2} in \App \ref{appendix-1}. For the type flag $s\in \set{0, 1}$ and $\forall n \in \mathbb{Z}_+$, let
	\begin{equation}
	\begin{aligned}
	G_s(n,\tau,\alpha_1, \alpha_2,\beta_1,\beta_2) 
	= \set{\binom{x}{y}\in \Omega: x\tau + y = n + s\tau} 
	\end{aligned}
	\end{equation}
be the set of the solutions to the Diophantine equation $x\tau + y = n + s\tau$ with rectangular constraint $\plpt{x}{y}\in \Omega$	. 
The cardinality of $G_s(n,\tau,\alpha_1, \alpha_2,\beta_1,\beta_2)$ is denoted by $\abs{G_s(n,\tau,\alpha_1, \alpha_2,\beta_1,\beta_2)}$. The geometric interpretation of the set $G_s(n,\tau,\alpha_1, \alpha_2,\beta_1,\beta_2)$ is all of the 2-dim points which satisfy \eqref{eq-xy-diop} with interger coordinates on the line $x\tau +y = n+s\tau$ and in the rectangle domain $\Omega$. \Fig \ref{fig-Dipahn-eq-geo} illustrates the scenario intuitively.

The equation $x\tau +y = n+s\tau$ implies that $y = n+s\tau - x\tau$. Thus $\beta_1\le y\le \beta_2$ implies that 
\begin{equation}
\frac{n+s\tau-\beta_2}{\tau} \le x \le \frac{n+s\tau-\beta_1}{\tau}
\end{equation}
It is obvious that
\begin{equation} \label{eq-Dioph-sol}
\max\left(\alpha_1, \frac{n+s\tau-\beta_2}{\tau}\right) \le x \le \min\left(\alpha_2, \frac{n+s\tau-\beta_1}{\tau}\right)
\end{equation}
For the purpose of finding integer solutions, the $\dfrac{n+s\tau-\beta_2}{\tau}$ should be replaced by $\mceil{\dfrac{n+s\tau-\beta_2}{\tau}}$ and the $\dfrac{n+s\tau-\beta_1}{\tau}$ should be replaced by $\mfloor{\dfrac{n+s\tau-\beta_1}{\tau}}$, where  $\mceil{x}$ and $\mfloor{x}$ denote the ceiling and floor of $x\in \mathbb{R}$ respectively. 
 Let
\begin{equation} \label{eq-Dioph-x-range}
\left\{
\begin{aligned}
\scrd{x}{min}^s(n,\tau, \alpha_1,\beta_2) &= \max\left(\alpha_1, \mceil{\frac{n+s\tau-\beta_2}{\tau}}\right)\\
\scrd{x}{max}^s(n,\tau,\alpha_2,\beta_1) &= \min\left(\alpha_2, \mfloor{\frac{n+s\tau-\beta_1}{\tau}}\right)
\end{aligned}
\right.
\end{equation}
for $s\in \set{0,1}$, then it is easy to prove the following lemma according to  \eqref{eq-Dioph-sol} and \eqref{eq-Dioph-x-range}. 

\begin{lem} \label{thm-sol-Diophantine}
For the constrained Diophantine equation 
\begin{equation} \label{eq-Dioph-c}
\tau x + y = n + s\tau, \quad \plpt{x}{y}\in \Omega, s\in \set{0,1}
\end{equation}
the set of its solutions can be written by
\begin{equation}
\begin{aligned}
G_s(n,\tau, \alpha_1, \alpha_2, \beta_1, \beta_2)
 =\set{\plpt{x}{n+s\tau-x\tau}: \scrd{x}{min}^s \le  x  \le \scrd{x}{max}^s}
\end{aligned}
\end{equation}
and the number of solutions is
\begin{equation}
\begin{aligned}
\abs{G_s(n,\tau, \alpha_1, \alpha_2, \beta_1, \beta_2)} 
= \scrd{x}{max}^s(n,\tau,\alpha_2,\beta_1) - \scrd{x}{min}^s(n,\tau, \alpha_1,\beta_2)+1.
\end{aligned}
\end{equation}
\end{lem}

Some special cases for the set $G_s(n,\tau, \alpha_1, \alpha_2, \beta_1, \beta_2)$ is discussed in \App \ref{app-B}.

\section{Bugs and Correction for the Key Formula in SGMD}

\subsection{Bugs in the SGMD Method}
The key formula of Pan et al. can be stated by the following  result \cite{Pan2019sgmd}:
\begin{res}	 \label{res-Pan}
	Suppose that $d$ is the embedding dimension, $m = N-(d-1)\tau$,  and $\mat{Z}_k = (Z^i_j(k))_{d\times m}\in \ES{R}{d}{m}$ is the $k$-th component of the trajectory matrix $\mat{M}=\Phi_{1,1}(X)$ in the immersion space $\ES{R}{d}{m}$ for the time series $X=\set{x[n]}^N_{n=1}$. Let
	\begin{equation}
	\tilde{Z}^i_j(k)  = \begin{cases}
	Z^i_j(k) , & m < d; \\
	Z^j_i(k) , & m \ge d.
	\end{cases}
	\end{equation}
for $1\le i\le d$ and $1\le j\le m$, the $k$-th component of time series $x^{(k)}[n]$ rebuilt from the $\mat{Z}_k$ can be computed by 
	\begin{equation}\label{eq-Pan} 
	\begin{aligned}
	x^{(k)}[n] 
	&= \begin{cases}  
	  \cfrac{1}{n}\sum\limits^{n}_{p=1}\tilde{Z}^p_{n-p+1}(k) , & 1 \le n < d^*;\\
	  \cfrac{1}{d^*}\sum\limits^{d^*}_{p=1} \tilde{Z}^p_{n-p+1}(k), & d^* \le n\le m^*; \\
	  \cfrac{1}{N-n+1}\sum\limits_{p=n-m^*+1}^{N-m^*+1}\tilde{Z}^p_{n-p+1}(k), & m^* < n \le N
	\end{cases}		
	\end{aligned}
	\end{equation}
where 
\begin{equation}
d^*=\min(d, m), \quad m^*=\max(d,m).
\end{equation}
\end{res}

We remark that the rebuilding process based on \eqref{eq-Pan} does not depend on the time delay $\tau$, which is conflict with the definition of the trajectory matrix $\scrd{\mat{X}}{sgmd}$ in \eqref{eq-M-pan} or its equivalent form $\mat{M}$ for $s=1$ in \eqref{eq-M-s-tau}. Consequently, the equation \eqref{eq-Pan} just holds for $\tau=1$. 
Moreover, the MATLAB code released by
Pan et al. can not be applied if $\tau>1$ or $s=0$.  

\subsection{Correction for the SGMD Method}

Our study shows that the bug in SGMD can be fixed if the DAP is generalized correctly.  We now propose the following pulling back theorem about the \textit{generalized diagonal averaging principle} (GDAP) for the SGMD and SSA:

\begin{thm}[Pulling Back Theorem] \label{thm-pullback-noisyfree}
For the type $s\in\set{0,1}$, embedding dimension $d\in \mathbb{N}$ and time series $X=\set{x[n]}^{N-1+s}_{n=s}$ of length$N$, suppose $\mat{M}=\Phi_{s,\tau}(X)\in \ES{R}{d}{m}$ is the trajectory matrix of $X$ with $r$ components $\mat{Z}_1, \cdots \mat{Z}_r$ such that $\displaystyle \mat{M} = \sum^r_{k=1}\mat{Z}_k$. Let
\begin{equation} \label{eq-qmin-qmax}
\left\{
\begin{aligned}
\scrd{q}{min} &= \max\left(s, \mceil{\frac{n -m +s\tau+ (1-s)}{\tau}} \right) \\
\scrd{q}{max} &= \min \left(d+s-1, \mfloor{\frac{n+(\tau-1)s}{\tau}}\right)
\end{aligned}
\right.
\end{equation}
and
\begin{equation} \label{eq-Q-nstau}
Q_k(n,\tau,s) =\set{Z^q_{n+s\tau -q\tau}(k): \scrd{q}{min}\le q\le \scrd{q}{max}}
\end{equation}
be the data set specified by the matrix $\mat{Z}_k=(Z^i_j(k))_{d\times m}\in \ES{R}{d}{m}$, then  the $k$-th component of time sequence $X^{(k)}=\set{x^{(k)}[n]}^{N-1+s}_{n=s}\in \mathscr{X}$ corresponding to the component $\mat{Z}_k$ can be computed by 
$$
X^{(k)} = \inv{\Phi}_{s,\tau}(\mat{Z}_k)=\set{x^{(k)}[n]: s\le n\le N-1+s}
$$
where
\begin{equation} \label{eq-xn-rebuild}
\begin{aligned}
x^{(k)}[n] 
&= \frac{1}{\abs{Q_k(n,\tau,s)}}\sum_{z\in Q_k(n,\tau,s)} z\\
&=\frac{1}{\scrd{q}{max} - \scrd{q}{min} + 1} \sum^{\scrd{q}{max}}_{q=\scrd{q}{min}}Z^q_{n+s\tau - q\tau}(k).
\end{aligned}
\end{equation}
for $1\le k\le r$.
\end{thm}

\noindent \textbf{Proof}: 

In order to find the inverse $X = \inv{\Phi}_{s,\tau}(\mat{M})$ such that the decomposition $\mat{M}=\sum^r_{k=1}\mat{Z}_k$, what we should do is to find the position of $x^{(k)}[n]$ appearing in the $k$-th component matrix $\mat{Z}_k$.  

According to the definition of the trajectory matrix $\mat{M}=\sum^r_{k=1}\mat{Z}_k$ of type-0 in \eqref{eq-X-traj-mat-1} or of type-1 in  \eqref{eq-X-traj-mat-0}, it is equivalent to find the solution $\plpt{i}{j}$ to the Diophantine equation
$\tau i + j = n$ or $\tau i + j = n +\tau$ for the given $n$. Obviously, the key issue lies in solving the binary Diophantine equation $\tau x + y = n + s\tau$ for $s\in \set{0, 1}$. According to the \Lem \ref{thm-sol-Diophantine}, the pair of row and column indices $\plpt{i}{j}$ are the elements of $G_s(n,\tau, \alpha_1, \alpha_2, \beta_1, \beta_2)$ for $(s, \alpha_1, \alpha_2, \beta_1, \beta_2) = (0, 0, d-1, 0, m-1)$ or $(s, \alpha_1, \alpha_2, \beta_1, \beta_2) = (1, 1, d, 1, m)$ with the form $\plpt{i}{j} = \plpt{q}{n+s\tau -q\tau}$ for $ \scrd{q}{min} \le q\le \scrd{q}{max}$.      

On the other hand, the data set 
\begin{equation}
\begin{aligned}
Q_k(n,\tau,0) 
= \set{Z^i_j(k): \plpt{i}{j}\in G_0(n,\tau,0,d-1,0,m-1)} 
=\set{Z^q_{n+s\tau-q\tau}(k): s=0, \scrd{q}{min}\le q\le \scrd{q}{max}}
\end{aligned}
\end{equation} 
for $s=0$ or 
\begin{equation}
\begin{aligned}
Q_k(n,\tau,1)
= \set{Z^i_j(k): \plpt{i}{j}\in G_1(n,\tau,1,d,1,m)} 
= \set{Z^q_{n+s\tau-q\tau}(k): s=1, \scrd{q}{min}\le q\le \scrd{q}{max}}
\end{aligned}
\end{equation} 
for $s=1$ contains all of the candidates or copies of $x^{(k)}[n]$ appearing in the matrix $\mat{Z}_k$. 
Consequently, the $x^{(k)}[n]$ can be rebuilt  by averaging all of the elements in the data set $Q_k(n,\tau,s)$, which is computed by \eqref{eq-xn-rebuild}. This completes the proof. $\blacksquare$

The \Thm \ref{thm-pullback-noisyfree} essentially states the method for computing the pulling back mapping $\inv{\Phi}_{s,\tau}: \ES{R}{d}{m}\to \mathscr{X}$ for computing the time sequence $X^{(k)}=\inv{\Phi}_{s,\tau}(\mat{Z}_k)$, thus it can be called the \textit{pulling back theorem} intuitively.

\subsection{Corollaries}

There are some trivial corollaries for $s\in\set{0,1}$ which can be applied for solving the components of the time sequence with concrete programming languages.

\subsubsection{Type-0 Time Series in C/C++/Pyhton/Rust/...}

For $s=0$ and $\tau\in \mathbb{N}$, we immediately have the following corollary by \Thm \ref{thm-pullback-noisyfree}:  
\begin{cor}	\label{cor-ZHY-C}
	For the $n\in \set{0, 1, 2, \cdots, N-1}$ and the $k$-th component $\mat{Z}_k = (Z^i_j(k))_{d\times m}\in \ES{R}{d}{m}$ of the trajectory matrix $\mat{Z}$  constructed from the time series $X=\set{x[n]}^{N-1}_{n=0}$ such that $\mat{Z}=\sum^r_{k=1}\mat{Z}_k=\Phi_{0,\tau}(X)$, let 
\begin{equation}
\left\{
\begin{aligned}
\scrd{p}{min} &= \max\left(0,\mceil{\frac{n-m+1}{\tau}} \right);  \\
\scrd{p}{max} &= \min\left(d-1, \mfloor{\frac{n}{\tau}}\right),
\end{aligned}
\right.
\end{equation}	
then we can convert the  $\mat{Z}_k$  to the
time series $X^{(k)}=\set{x^{(k)}[n]}^{N-1}_{n=0} =\inv{\Phi}_{0,\tau}(\mat{Z}_k)$ by 
	\begin{equation}
	x^{(k)}[n] = \frac{1}{\scrd{p}{max}-\scrd{p}{min}+1}\sum^{\scrd{p}{max}}_{p=\scrd{p}{min}}Z^{p}_{n-p\tau}(k).
	\end{equation}	
\end{cor}

\subsubsection{Type-1 Time Series in MATLAB/Octave/Fortran}

For $s=1$ and $\tau\in \mathbb{N}$, we immediately have the following corollary by \Thm \ref{thm-pullback-noisyfree}:  
\begin{cor}	\label{cor-ZHY-fortran}
	For the $n\in \set{1, 2, \cdots, N}$ and the $k$-th component $\mat{Z}_k = (Z^i_j(k))_{d\times m}\in \ES{R}{d}{m}$ of the trajectory matrix $\mat{Z}$  constructed from the time series $X=\set{x[n]}^{N}_{n=1}$ such that $\displaystyle\mat{Z}=\sum^r_{k=1}\mat{Z}_k=\Phi_{1,\tau}(X)$, let 
\begin{equation}
\left\{
\begin{aligned}
\scrd{p}{min} &= \max\left(1, \mceil{\frac{n+\tau-m}{\tau}}\right)  \\
\scrd{p}{max} &= \min\left(d,\mfloor{\frac{n+\tau-1}{\tau}}\right)
\end{aligned}
\right.
\end{equation}	
then we can convert the  $\mat{Z}_k$  to the
time series $X^{(k)}=\set{x^{(k)}[n]}^{N}_{n=1} =\inv{\Phi}_{1,\tau}(\mat{Z}_k)$ by 
	\begin{equation}
	x^{(k)}[n] = \frac{1}{\scrd{p}{max}-\scrd{p}{min}+1}\sum^{\scrd{p}{max}}_{p=\scrd{p}{min}}Z^{p}_{n+\tau-p\tau}(k).
	\end{equation}	
\end{cor}
The \textbf{Corollary} \ref{cor-ZHY-fortran} is the correct version for rebuilding the components of time series from the components of trajectory matrix, which depends on the time delay $\tau$ explicitly and can be applied for the type-1 time series.

\subsubsection{Equivalent Form of DAP in SGMD and SSA subjecto to $(s,\tau)=(1,1)$ }
 
For $s=1$ and $\tau=1$, we immediately have the following corollary by \Thm \ref{thm-pullback-noisyfree}:  
\begin{cor}	\label{cor-ZHY-matlab}
	For the $n\in \set{1, 2, \cdots, N}$ and the $k$-th component $\mat{Z}_k = (Z^i_j(k))_{d\times m}\in \ES{R}{d}{m}$ of the trajectory matrix $\mat{Z}$  constructed from the time series $X=\set{x[n]}^{N}_{n=1}$ such that $\mat{Z}=\sum^r_{k=1}\mat{Z}_k=\Phi_{1,1}(X)$, let
\begin{equation}
\left\{
\begin{aligned}
\scrd{p}{min} &= \max\left(1, n+1-m\right);  \\
\scrd{p}{max} &= \min\left(d,n\right),
\end{aligned}
\right.
\end{equation}	
then we can convert the  $\mat{Z}_k$  to the
time series $X^{(k)}=\set{x^{(k)}[n]}^{N}_{n=1} =\inv{\Phi}_{1,1}(\mat{Z}_k)$ by 
	\begin{equation}
	x^{(k)}[n] = \frac{1}{\scrd{p}{max}-\scrd{p}{min}+1}\sum^{\scrd{p}{max}}_{p=\scrd{p}{min}}Z^{p}_{n+1-p}(k).
	\end{equation}	
\end{cor}

We remark that \Cor \ref{cor-ZHY-matlab} is equivalent to the \textbf{Result} \ref{res-Pan} according to the equivalent form of $G_1(n,1,1,d,1,m)$ proposed in \App \ref{app-B.3}, where an intuitive geometric interpretation is illustrated in \Fig \ref{fig-G1}.

\section{Conclusion} \label{sect-conclusions}

The original SGMD  method should be used carefully since the DAP are limited by two rules: 
the time delay $\tau$ must be set with $\tau=1$; and the time series must be of type-1 with the form $X=\set{x[n]}^N_{n=1}$.

For the time delay $\tau>1$ and time series $X = \set{x[n]}^{N-1+s}_{n=s}$ of type-$s$ and length $N$, the components of time series should be computed according to \eqref{eq-xn-rebuild} for $s\in \set{0,1}$.  Our new formula holds for time series denoted by $X=\set{x[n]}^{N-1+s}_{n=s}$ and any time delay $\tau\ge 1$, which can be applied not only for the SGMD method but also for the SSA method.

\subsection*{Acknowledgments}

This work was supported in part by the National Natural Science Foundation of China under grant numbers 62167003 and 62373042, in part by the Hainan Provincial Natural Science Foundation of China under grant number 720RC616, in part by the Research Project on Education and Teaching Reform in Higher Education System of Hainan Province under grant number Hnjg2025ZD-28, in part by the Specific Research Fund of the Innovation Platform for Academicians of Hainan Province, in part by the Guangdong Basic and Applied Basic Research Foundation under grant number 2023A1515010275, and in part by the Foundation of National Key Laboratory of Human Factors Engineering under grant number HFNKL2023WW11, in part by the Hainan Province Key R \& D Program Project under grant number WSJK2024MS234.

\subsection*{Data Availability}

Not applicable

\subsection*{Code Availability}

Not applicable

\subsection*{Declaration of interests}

 The authors declare that they have no known competing financial interests or personal relationships that could have appeared to influence the work reported in this paper.
 
\begin{appendix}

\section{Diophantine Equation $\tau x + y =n + s\tau $} \label{appendix-1}
For the Diophantine equation $a x + by =n$, we have the following important properties \cite{HuaLK1958,Cohen2007GTM239}: 

\begin{thm} \label{thm-appendix-1}
For $a, b\in \mathbb{N}=\set{1,2,3,\cdots}$ and $n\in \mathbb{Z}=\set{0, \pm 1, \pm2, \cdots}$, the equation $ax+by = n$ has solution $\plpt{x}{y}\in \mathbb{Z}^2$ if and only if the greatest common divisor of $a$ and $b$ is a factor of $n$, i.e., $\GCD(a,b)\mid n$.
\end{thm}
\begin{thm} \label{thm-appendix-2}
Suppose that $a, b\in \mathbb{N}$ such that $\GCD(a,b) = 1$. For any $n\in \mathbb{Z}$ such that $n> ab -a-b$, there must exist $x, y\in \mathbb{Z}^+ = \set{0, 1, 2, \cdots}$ such that $n = ax + by$.
\end{thm}

If we take the assignments for the parameters $a$, $b$ and $s$ as
follows
\begin{equation}
a \gets \tau, b \gets 1, n \gets n+s\tau
\end{equation}
for $s\in \set{0,1}$ and $\tau\in \mathbb{N}$, then the solution of the Diophantine equation $\tau x + y = n + s\tau$ have non-negative solution $\plpt{x}{y}\in (\mathbb{Z}^+)^2$ for any $n \in \mathbb{Z}^+$.

\section{Special Cases for $G_s(n,\tau,\alpha_1, \alpha_2,\beta_1,\beta_2)$} \label{app-B}

There are some typical special cases about the structure of $G_s(n,\tau, \alpha_1, \alpha_2, \beta_1, \beta_2)$ for $s\in \set{0, 1}$, $(\alpha_1, \alpha_2) \in \set{(0, d-1), (1, d)}$ and $(\beta_1, \beta_2)\in \set{(0, m-1), (1, m)}$  because of the way of representation of time sequences and coding with concrete computer programming languages. 

\subsection{Case of $(\alpha_1, \alpha_2, \beta_1, \beta_2) = (0, d-1, 0, m-1)$} \label{app-B.1}

Particularly, for $d, m\in \mathbb{N}$ and  $(\alpha_1, \alpha_2, \beta_1, \beta_2) = (0, d-1, 0, m-1)$, we have
\begin{equation} \label{eq-set-G0-boundary}
\left\{
\begin{aligned}
\scrud{x}{0}{min}(n,\tau, 0,m-1) & =
\max\left(0,\mfloor{\frac{n-m+1}{\tau}}\right) \\
\scrud{x}{0}{max}(n,\tau, d-1,0) & =
\min\left(m-1,\mfloor{\frac{n}{\tau}}\right) 
\end{aligned}
\right.
\end{equation}
and 
\begin{equation} \label{eq-set-G0}
\begin{aligned}
G_0(n,\tau, 0, d-1, 0, m-1) 
= \left\{\plpt{x}{n-x\tau}: \scrud{x}{0}{min} \le x \le \scrud{x}{0}{max} \right\} 
\end{aligned}
\end{equation}

\subsection{Case of $(\alpha_2,\alpha_2, \beta_1, \beta_2) = (1, d, 1, m)$} \label{app-B.2}
 
Particularly, for $d, m\in \mathbb{N}$ and  $(\alpha_1, \alpha_2, \beta_1, \beta_2) = (1, d, 1, m)$, we can find that 
\begin{equation} \label{eq-set-G1-boundary}
\left\{
\begin{aligned}
\scrud{x}{1}{min}(n,\tau, 1, m) &= \max\left(1, \mceil{\frac{n+\tau-m}{\tau}} \right)\\
\scrud{x}{1}{max}(n,\tau, d,1)  &= \min\left(d, \mfloor{\frac{n+\tau -1}{\tau}} \right)
\end{aligned}
\right.
\end{equation}
and
\begin{equation} \label{eq-set-G1}
\begin{aligned}
G_1(n,\tau, 1, d, 1, m) 
= \set{\plpt{x}{n+\tau-x\tau}:  \scrud{x}{1}{min}\le x   \le   \scrud{x}{1}{max}} 
\end{aligned}
\end{equation}

\subsection{Case of $(\alpha_1,\alpha_2,\beta_1,\beta_2)=(1,d,1, m)$ and  $\tau=1$} \label{app-B.3}

\begin{figure*}[hbtp]
\centering
\includegraphics[width=0.75\textwidth]{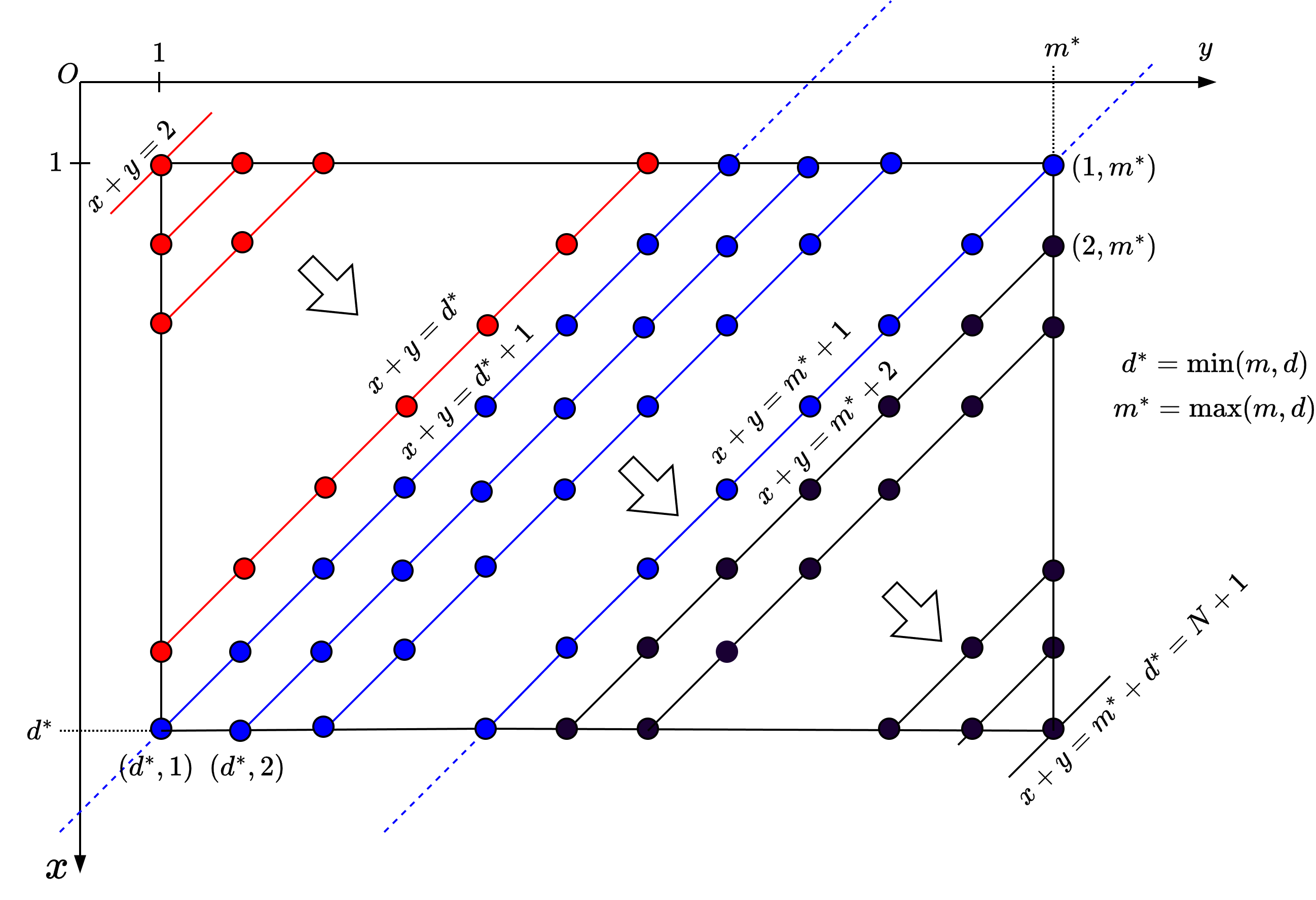} 
\caption{Geometric interpretation of $G_1(n, 1, 1, d, 1, m)$ in the DAP proposed in SSA} \label{fig-G1}
\end{figure*} 

A special case for the $G_1(n,\tau,\alpha_1,\alpha_2,\beta_1,\beta_2)$ such that $(\tau,\alpha_1,\alpha_2,\beta_1,\beta_2) = (1,1,d,1,m)$ is of interest in the references. \Fig  \ref{fig-G1} illustrates such a scenario.  
Let
\begin{equation}
d^* = \min(d, m), \quad m^* = \max(d, m)
\end{equation}
then $d + m = \min(d,m) + \max(d, m)=d^*+m^*$. Thus $N = m + (d-1)\tau = m + d - 1 = d^* + m^* -1$ or equivalently $d^*+m^*=N+1$. For the set
\begin{equation}
\begin{aligned}
G_1^*(n) 
= G_1(n,1,1,d,1,m) 
= \set{\plpt{x}{n+1-x}: \scrud{x}{1}{min}\le x\le \scrud{x}{1}{max}},
\end{aligned}
\end{equation}
we can obtain some interesting observations from \Fig  \ref{fig-G1}:
\begin{itemize}
\item[\ding{172}] For $1\le n< d^*$, there are $n$ solutions to the Diophantine equation $\tau x + y = n + \tau $ for $\plpt{x}{y}\in \Omega(1,d,1,m)$ and $s=1$, thus
\begin{equation} \label{eq-G1-1}
\left\{
\begin{aligned}
&G_1^*(n) 
= \left\{\plpt{x}{n+1-x}: 1\le x\le n\right\} \\
&\abs{G_1^*(n)} = n
\end{aligned}
\right. 
\end{equation}
Geometrically, the $G_1^*(n)$ corresponds to the red points on the red hypotenuse of the equilateral right angled triangle  shown in the \Fig  \ref{fig-G1}.
\item[\ding{173}] For $d^* \le n \le m^*$, there are always $d^*$ solutions to  the   Diophantine equation
$\tau x + y = n + s\tau$ for $\plpt{x}{y}\in \Omega(1,d,1,m)$, thus
\begin{equation} \label{eq-G1-2}
\left\{
\begin{aligned}
G_1^*(n) 
&= \set{\plpt{x}{n+1-x}: 1+ \max(0,n-m)\le x  \le \min(d,n)},  \\
\abs{G_1^*(n)} &= d^*
\end{aligned}
\right. 
\end{equation} 
Geometrically, the $G_1^*(n)$ corresponds to the blue points on the blue edge of the parallelogram  shown in the \Fig  \ref{fig-G1}.
\item[\ding{174}] For $m^* < n \le N$, there are $N-n+1$ solutions to  the  Diophantine equation
$\tau x + y = n + s\tau$ for $\plpt{x}{y}\in \Omega(1,d,1,m)$, thus
\begin{equation} \label{eq-G1-3}
\left\{
\begin{aligned}
G_1^*(n) 
&= \set{\plpt{n+1-y}{y}: n-m^* +1\le y  \le N-m^* +1}  \\
&= \set{\plpt{x}{n+1-x}: m^* +n - N \le x \le m^*}\\
\abs{G_1^*(n)} &= N-n+1
\end{aligned}
\right. 
\end{equation} 
for $m^* < n\le N$.
Geometrically, the $G_1^*(n)$ corresponds to the black points on the black edge of the parallelogram  shown in the \Fig  \ref{fig-G1}.
\end{itemize}

\section{Example of Embedding and Pulling Back}

For the purpose of illustrating the scenarios of embedding and pulling back mapping, we set $(N, d, m, \tau) = (27, 7, 9, 3)$ for the time series $X$ and observe the the trajectory matrix and the set $G_s(n, \tau, \alpha_1,\alpha_2,\beta_1,\beta_2)$ for $s\in\set{0,1}$. 

\subsection{Type-0 Time Series}

For $s=0$ and $(N, d, m, \tau) = (27, 7, 9, 3)$, the time series must be
$$
X=\set{x[0], x[1], \cdots, x[26]}.
$$
Its trajectory matrix is shown in \Fig \ref{fig-X-trajmat-python}.
 \begin{figure*}[htbp]
 \centering
 $$
\begin{bmatrix}
 x[0] & x[1] & x[2] & x[3] & x[4]& x[5] & x[6] & x[7]& x[8] \\
 x[3] & x[4] & x[5] & x[6] & x[7] & x[8] & x[9] & x[10] & x[11] \\
 x[6] & x[7] & x[8] & x[9] & x[10] & x[11] &x[12] & x[13] & x[14]  \\
 x[9] & x[10] & x[11] &x[12] & x[13] & [14] &x[15] & x[16] & x[17]   \\
 x[12] & x[13] & [14] &x[15] & x[16] & x[17] & x[18] & x[19] & x[20]  \\
 x[15] & x[16] & x[17] & x[18] & x[19] & x[20] & x[21] & x[22] & x[23] \\
 x[18] & x[19] & x[20] & x[21] & x[22] & x[23] & x[24] & x[25] & x[26] \\
 \end{bmatrix}
 $$
 \caption{Convert $X=\set{x[n]}^{26}_{n=0}$ to the trajectory matrix $\mat{M}=(M^i_j)\in \ES{R}{7}{9}$ with $\tau= 3$ for $0\le i\le 6$ and $ 0\le j\le 8$ }
 \label{fig-X-trajmat-python}
 \end{figure*}

On the other hand, we can  set $(\alpha_2,\alpha_2, \beta_1, \beta_2) = (0, 6, 0, 8)$ for the domain $\Omega$. Equation \eqref{eq-set-G0-boundary} implies that
\begin{equation}
\left\{
\begin{aligned}
&\scrud{x}{0}{min}(n,3,0,8) = \max\left(0, \mceil{\frac{n-8}{3}}\right) \\
&\scrud{x}{0}{max}(n,3,6,0) = \min\left(6, \mfloor{\frac{n}{3}}\right) \\
\end{aligned}
\right.
\end{equation}
and equation \eqref{eq-set-G0} implies that
\begin{equation}
\begin{aligned}
G_0(n,3,0,6,0,8) 
= \set{\plpt{x}{n-3x}: \scrud{x}{0}{min}(n,3,0,8) \le x \le \scrud{x}{0}{max}(n,3,6,0)}
\end{aligned}
\end{equation}
The structure of $G_0(n,\tau, 0, d-1, 0, m-1)= G_0(n,3,0, 6, 0, 8)$ is listed in \Tab \ref{tab-Dioph-eg-sol}. 

\begin{table}[H]
\centering
\caption{Structure of $G_0(n,3,0,6,0,8)=\set{\plpt{x}{n-3x}:  \scrud{x}{0}{min} \le x \le \scrud{x}{0}{max}}$ for $\scrud{x}{0}{min}= \max\left(0, \mceil{\frac{n-8}{3}}\right)$ and $\scrud{x}{0}{max}=\min\left(6, \mfloor{\frac{n}{3}}\right)$} \label{tab-Dioph-eg-sol}
\begin{tabular}{ccclc}
\hline
$n$ & $\scrud{x}{0}{min}$ & $\scrud{x}{0}{max}$  & $G_0(n,3,0,6,0,8)$  &  $\abs{G_0(n,3,0,6,0,8)}$  \\
\hline
 $0$ & $0$ & $0$ & $\set{\plpt{0}{0}}$  & $1$ \\
$1$ & $0$ &  $0$ & $\set{\plpt{0}{1}}$  & $1$ \\
$2$ & $0$ & $0$ & $\set{\plpt{0}{2}}$  & $1$\\
$3$ & $0$ & $1$ & $\set{\plpt{0}{3}, \plpt{1}{0}}$ & $2$\\
$4$ & $0$ & $1$ & $\set{\plpt{0}{4}, \plpt{1}{1}}$  & $2$ \\
$5$ & $0$ & $1$ & $\set{\plpt{0}{5}, \plpt{1}{2}}$  & $2$ \\
$6$ & $0$ & $2$ & $\set{\plpt{0}{5}, \plpt{1}{2}, \plpt{2}{0}}$  & $3$ \\
$7$ & $0$ & $2$ & $\set{\plpt{0}{7}, \plpt{1}{4}, \plpt{2}{1}}$  & $3$ \\
$8$ & $0$ & $2$ & $\set{\plpt{0}{8}, \plpt{1}{5}, \plpt{2}{2}}$  & $3$ \\
$9$ & $1$ & $3$ & $\set{\plpt{1}{6}, \plpt{2}{3}, \plpt{3}{0}}$  & $3$ \\
$10$& $1$ & $3$ & $\set{\plpt{1}{7}, \plpt{2}{4}, \plpt{3}{1}}$  & $3$ \\
$11$& $1$ & $3$ & $\set{\plpt{1}{8}, \plpt{2}{5}, \plpt{3}{2}}$  & $3$ \\
$12$& $2$ & $4$ & $\set{\plpt{2}{6}, \plpt{3}{3}, \plpt{4}{0}}$  & $3$ \\ 
$13$& $2$ & $4$ & $\set{\plpt{2}{7}, \plpt{3}{4}, \plpt{4}{1}}$  & $3$ \\
$14$& $2$ & $4$ & $\set{\plpt{2}{8}, \plpt{3}{5}, \plpt{4}{2}}$  & $3$ \\
$15$ & $3$ & $5$ & $\set{\plpt{3}{6}, \plpt{4}{3}, \plpt{5}{0}}$ & $3$\\
 $\vdots$ &  $\vdots$ &  $\vdots$ & \quad $\vdots$  & $\vdots$ \\
$26$ & $6$& $6 $  & $\set{\plpt{6}{8}}$ & $1$\\
\hline
\end{tabular}
\end{table}

\Tab \ref{tab-Dioph-eg-sol} describes the position of $x[n]$ in the trajectory matrix $\mat{M}$, which can be used to estimate the $x[n]$ with generalized diagonal averaging principle.  For example, when $n=9$, we have 
$$G_0(9,3,0,6,0,8) = \set{\plpt{1}{6}, \plpt{2}{3}, \plpt{3}{0}},$$ which means $x[9]$ appears as the entries $M^1_6, M^2_3$ and $M^3_0$ in the trajectory matrix $\mat{M}$ shown in \Fig \ref{fig-X-trajmat-python}.
In other words, we can take the pulling back theorem about the GDAP to rebuild the $x[9]$ by $x[9] = \frac{1}{3}(M^1_6 + M^2_3 + M^3_0)$.

\subsection{Type-1 Time Series}

For $s=1$ and $(N, d, m, \tau) = (27, 7, 9, 3)$, the time series must be
$$
X=\set{x[1], x[2], \cdots, x[27]}.
$$
Its trajectory matrix is shown in \Fig \ref{fig-X-trajmat-matlab}.

 \begin{figure*}[htbp]
 \centering
$$
 \begin{bmatrix}
 x[1] & x[2] & x[3] & x[4]& x[5] & x[6] & x[7]& x[8] & x[9] \\
 x[4] & x[5] & x[6] & x[7] & x[8] & x[9] & x[10] & x[11] & x[12]\\
 x[7] & x[8] & x[9] & x[10] & x[11] &x[12] & x[13] & x[14] & x[15] \\
 x[10] & x[11] &x[12] & x[13] & [14] &x[15] & x[16] & x[17] & x[18]  \\
 x[13] & [14] &x[15] & x[16] & x[17] & x[18] & x[19] & x[20]  & x[21]\\
 x[16] & x[17] & x[18] & x[19] & x[20] & x[21] & x[22] & x[23] & x[24]\\
 x[19] & x[20] & x[21] & x[22] & x[23] & x[24] & x[25] & x[26] & x[27] \\
 \end{bmatrix}
 $$
 \caption{Convert $X=\set{x[n]}^{27}_{n=1}$ to the trajectory matrix $\mat{M}=(M^i_j)\in \ES{R}{7}{9}$ with $\tau= 3$ for $1\le i\le 7$ and $ 1\le j\le 9$  }
 \label{fig-X-trajmat-matlab}
 \end{figure*}
 
 On the other hand, we can  set $(\alpha_2,\alpha_2, \beta_1, \beta_2) = (1, 7, 1, 9)$ for the domain $\Omega$. With the help of \eqref{eq-set-G1-boundary} and \eqref{eq-set-G1}, we immediately have 
\begin{equation}
\left\{
\begin{aligned}
&\scrud{x}{1}{min}(n,3,1,9) = \max\left(1, \mceil{\frac{n-6}{3}}\right) \\
&\scrud{x}{1}{max}(n,3,7,1) = \min\left(7, \mfloor{\frac{n+2}{3}}\right) \\
&G_1(n,3,1,7,1,9) 
= \set{\plpt{x}{n+3-3x}: \scrud{x}{1}{min} \le x \le \scrud{x}{1}{max}}
\end{aligned}
\right.
\end{equation}
The structure of $G_1(n,\tau, 1, d, 1, m)= G_1(n,3, 1, 7, 1, 9)$ is listed in \Tab \ref{tab-Dioph-eg-sol-2}.
\begin{table}[htb]
\centering
\caption{Structure of $G_1(n,3,1,7,1,9)=\tiny{\set{\plpt{x}{n+3-3x}:  \tiny{\scrud{x}{1}{min} \le x \le \scrud{x}{1}{max}}}}$ for $\scrud{x}{1}{min}= \max\left(1, \mceil{\frac{n-6}{3}}\right)$ and $\scrud{x}{1}{max}=\min\left(7, \mfloor{\frac{n+2}{3}}\right)$.} \label{tab-Dioph-eg-sol-2}
\begin{tabular}{ccclc}
\hline
$n$ & $\scrud{x}{1}{min}$ & $\scrud{x}{1}{max}$  & $G_1(n,3,1,7,1,9)$  &  $\abs{G_1(n,3,1,7,1,9)}$  \\
\hline
 $1$ & $1$ & $1$ & $\set{\plpt{1}{1}}$  & $1$ \\
$2$ & $1$ &  $1$ & $\set{\plpt{1}{2}}$  & $1$ \\
$3$ & $1$ & $1$ & $\set{\plpt{1}{3}}$ & $1$\\
$4$ & $1$ & $2$ & $\set{\plpt{1}{4}, \plpt{2}{1}}$  & $2$ \\
$5$ & $1$ & $2$ & $\set{\plpt{1}{5}, \plpt{2}{2}}$  & $2$ \\
$6$ & $1$ & $2$ & $\set{\plpt{1}{6}, \plpt{2}{3}}$  & $2$ \\
$7$ & $1$ & $3$ & $\set{\plpt{1}{7}, \plpt{2}{4}, \plpt{3}{1}}$  & $3$ \\
$8$ & $1$ & $3$ & $\set{\plpt{1}{8}, \plpt{2}{5}, \plpt{3}{2}}$  & $3$ \\
$9$ & $1$ & $3$ & $\set{\plpt{1}{9}, \plpt{2}{6}, \plpt{3}{3}}$  & $3$ \\
$10$& $2$ & $4$ & $\set{\plpt{2}{7}, \plpt{3}{4}, \plpt{4}{1}}$  & $3$ \\
$11$& $2$ & $4$ & $\set{\plpt{2}{8}, \plpt{3}{5}, \plpt{4}{2}}$  & $3$ \\
$12$& $2$ & $4$ & $\set{\plpt{2}{9}, \plpt{3}{6}, \plpt{4}{3}}$  & $3$ \\ 
$13$& $3$ & $5$ & $\set{\plpt{3}{7}, \plpt{4}{4}, \plpt{5}{1}}$  & $3$ \\
$14$& $3$ & $5$ & $\set{\plpt{3}{8}, \plpt{4}{5}, \plpt{5}{2}}$  & $3$ \\
$15$ & $3$ & $5$ & $\set{\plpt{3}{9}, \plpt{4}{6}, \plpt{5}{3}}$ & $3$\\
$16$ & $4$ & $6$ & $\set{\plpt{4}{7}, \plpt{5}{4}, \plpt{6}{1}}$ & $3$\\
 $\vdots$ & $\vdots$ & $\vdots$ & \quad $\vdots$  & $\vdots$ \\
$27$ & $7$& $7 $  & $\set{\plpt{7}{9}}$ & $1$\\
\hline
\end{tabular}
\end{table}

Similarly, \Tab \ref{tab-Dioph-eg-sol-2} describes the position of $x[n]\in X=\set{x[1], \cdots, x[27]}$ in the trajectory matrix $\mat{M}$.  For example, when $n=9$, we have 
$$G_1(9,3,1,7,1,9) = \set{\plpt{1}{9}, \plpt{2}{6}, \plpt{3}{3}},$$ which means $x[9]$ appears as the entries $M^1_9, M^2_6$ and $M^3_3$ in the trajectory matrix $\mat{M}$ shown in \Fig \ref{fig-X-trajmat-matlab}. The pulling back theorem about the GDAP implies  that the $x[9]$ can be estimated by $x[9] = \frac{1}{3}(M^1_9 + M^2_6 + M^3_3)$. 

\end{appendix}


\begin{thebibliography}{10}

\bibitem{Pan2019sgmd}
Haiyang Pan, Yu~Yang, Xin Li, Jinde Zheng, and Junsheng Cheng.
\newblock Symplectic geometry mode decomposition and its application to
  rotating machinery compound fault diagnosis.
\newblock {\em Mechanical Systems and Signal Processing}, 114(1):189--211,
  2019.
\newblock \url{https://doi.org/10.1016/j.ymssp.2018.05.019}.

\bibitem{Vautard1992SSA}
Robert Vautard, Pascal Yiou, and Michael Ghil.
\newblock Singular-spectrum analysis: A toolkit for short, noisy chaotic
  signals.
\newblock {\em Physica D: Nonlinear Phenomena}, 58(1):95--126, 1992.

\bibitem{Jin2019}
Hang Jin, jianhui Lin, Xieqi Chen, and Cai Yi.
\newblock Modal parameters identification method based on symplectic geometry
  model decomposition.
\newblock {\em Shock and Vibration}, 2019:5018732, 2019.
\newblock \url{ https://doi.org/10.1155/2019/5018732}.

\bibitem{ZhangGY2022Esgmd}
Guangyao Zhang, Yi~Wang, Xiaomeng Li, Baoping Tang, and Yi~Qin.
\newblock Enhanced symplectic geometry mode decomposition and its application
  to rotating machinery fault diagnosis under variable speed conditions.
\newblock {\em Mechanical Systems and Signal Processing}, 170:108841, 2022.
\newblock \url{https://doi.org/10.1016/j.ymssp.2022.108841}.

\bibitem{GuoJC2022}
Jianchun Guo, Zetian Si, Yi~Liu, Jiahao Li, Yanting Li, and Jiawei Xiang.
\newblock Dynamic time warping using graph similarity guided symplectic
  geometry mode decomposition to detect bearing faults.
\newblock {\em Reliability Engineering \& System Safety}, 224:108533, 2022.

\bibitem{GuoJC2023}
Jianchun Guo, Zetian Si, and Jiawei Xiang.
\newblock Cycle kurtosis entropy guided symplectic geometry mode decomposition
  for detecting faults in rotating machinery.
\newblock {\em ISA Transactions}, 138:546--561, 2023.

\bibitem{ChenYJ2023}
Yijie Chen, Zhenwei Guo, and Dawei Gao.
\newblock Marine controlled-source electromagnetic data denoising method using
  symplectic geometry mode decomposition.
\newblock {\em Journal of Marine Science and Engineering}, 11(8):1578, 2023.
\newblock \url{https://www.mdpi.com/2077-1312/11/8/1578}.

\bibitem{Liu2024SpSparse}
Yanfei Liu, Junsheng Cheng, Yu~Yang, Jinde Zheng, Haiyang Pan, Xingkai Yang,
  Guangfu Bin, and Yiping Shen.
\newblock Symplectic sparsest mode decomposition and its application in rolling
  bearing fault diagnosis.
\newblock {\em IEEE Sensors Journal}, 24(8):12756--12769, 2024.
\newblock \url{https://doi.org/10.1109/JSEN.2024.3370959}.

\bibitem{Hao2024}
Jingtang Hao, Long Ma, Xutao Yin, Xinyi Zhao, and Zhigang Su.
\newblock Improved symplectic geometry mode decomposition based correlation
  method in white light scanning interferometry.
\newblock {\em Optics and Lasers in Engineering}, 182:108482, 2024.
\newblock \url{https://doi.org/10.1016/j.optlaseng.2024.108482}.

\bibitem{ZhanPM2024}
Pengming Zhan, Xianrong Qin, Qing Zhang, and Yuantao Sun.
\newblock Output-only modal identification based on auto-regressive
  spectrum-guided symplectic geometry mode decomposition.
\newblock {\em Journal of Vibration Engineering \& Technologies},
  12(1):139--161, 2024.
\newblock \url{https://doi.org/10.1007/s42417-022-00832-1}.

\bibitem{XinG2025Csgmd}
Ge~Xin, Yifei Chen, Lingfeng Li, Chuanhai Chen, Zhifeng Liu, and Jérôme
  Antoni.
\newblock Complex symplectic geometry mode decomposition and a novel
  time–frequency fault feature extraction method.
\newblock {\em IEEE Transactions on Instrumentation and Measurement},
  74(1):1--10, 2025.

\bibitem{HuaLK1958}
Loo~Keng Hua.
\newblock {\em Introduction to Number Theory}.
\newblock Springer-Verlag, New York, 1982.
\newblock the Chinese version was published by the Science Press in 1957.

\bibitem{Cohen2007GTM239}
Henri Cohen.
\newblock {\em Number Theory, Volume I: Tools and Diophantine Equations},
  volume 239 of {\em Graduate Texts in Mathematics}.
\newblock Springer, New York, 2007.

\end{thebibliography}

\end{document}